\documentstyle[aps,prb,multicol,epsf,picinpar]{revtex}

\def\e2 {\epsilon-\epsilon_k}
\def\be {\begin{equation}}
\def\ee {\end{equation}}
\def\bea {\begin{eqnarray}}
\def\eea {\end{eqnarray}}
\def\om {\omega}

\begin{document}
\draft
\title{Dominant mode in the cuprates: electronic vs. phononic scenario
}
\author{George Kastrinakis}
\address{Institute for Electronic Structure and Laser (IESL), Foundation for
Research and Technology - Hellas (FORTH), \\
P.O. Box 1527, Iraklio, Crete 71110, Greece $^*$}

\date{February 4, 2002}
\maketitle
\begin{abstract}

We make the case for an electronic origin of the strong mode, recently
seen in ARPES experiments, and, in this regard, 
further discuss the physics of the spin resonance peak.

\end{abstract}

\vspace{0.7cm}

Recent ARPES experiments observe a 'kink' in the energy vs. momentum 
dispersion, in a number of cuprates - e.g. see \cite{lan,slin,bog}. 
The effect is more intense in
the superconducting state, but clearly persists in the normal state
as well. Moreover, optical conductivity experiments indicate a
matching reduction of the scattering rate \cite{tim}. 

It is well known that a sharp collective mode can generate an electronic
response \cite{eng,nor} in qualitative agreement with the aforementioned
data. The nature of this sharp mode, i.e. electronic, phononic etc. origin,
does not influence qualitatively the effect.

Refs. \cite{lan,slin} take the viewpoint that the dominant mode coupled 
to the carriers is of phononic origin. Below we argue for an electronic 
dominant mode picture.

As stated in \cite{bog,slin}, the well known spin resonance peak 
could, in principle, be interpreted as the sharp mode.
It is dismissed, afterall, on the basis that, {\em usually}, the 
spin resonance peak is not seen in the normal state.

In ref. \cite{gk} we have presented a model for the spin resonance peak,
which can consistently account for its appearance in the {\em normal}
state of Zn-doped YBCO \cite{fong}. The central idea here is that the 
spin resonance peak  - or some equivalent mode - is a many-body effect, present for {\em all} temperatures, and for a fairly broad range of parameters. 
The peak just becomes {\em sharper} in the superconducting state,
once the characteristic energy scale $\om_{res} < 2 \Delta$, where
$\Delta$ is the (maximum) gap - e.g. c.f. \cite{nor}. This fact can explain 
why the peak is visible by neutron scattering only in the SC state
of the pure YBCO and BSCCO. Both these materials are bilayers, and in
ref.  \cite{gk} a specific bilayer (easily extendable to multi-layer: e.g.
for a tri-layer system, {\em two} resonance peaks could appear)
model is proposed. However, the dominant peak in the susceptibility
of the carriers is of course present even for a monolayer system.
In general, a non-parabolic dispersion, such as the $t,t',t'',...$ used
for the cuprates, generates peaks in the susceptibility for various
momenta and energies, as a function of the filling factor and the
coupling - e.g. c.f. \cite{pao,gk}. These peaks can be both strong
and narrow. In the frame of a self-consistent Hubbard model calculation,
we obtain values for $\omega_{res}$ in the range $t/5$ to $t/13$ for 
$U \sim 4t - 6t$.

Besides the purely electronic contribution, the peak(s) can
have a magnetic component according to the model in \cite{gk}, which
can make the peaks even better defined. 
As mentioned in \cite{gk}, Zn-doping does increase the strength of
the susceptibility peak, due to the AF correlations enhancement.
We expect that Zn-doping can make observable the peak in materials
where it is only seen in the SC state, in the same manner as in YBCO.
Actually, in principle it is possible that Zn-doping may enhance 
the peak to the point of it becoming detectable even for materials such
as LSCO, in which it is not detectable even in the SC state.

There are two very recent expts. agreeing with the above. First, in
\cite{wang} it was shown that the famous ARPES feature is not of
phononic origin (but of rather electronic), via reflectivity measurements.
Second, in \cite{he} the monolayer Tl-2201 was shown to exhibit the
spin resonance peak in the SC state.

Thus it appears that the 
electronic (possibly enhanced by magnetism) scenario provides a viable 
explanation for the 'kink' feature seen in ARPES, in agreement with the
spin resonance peak observations.

\vspace{.5cm}
 $^*$ e-mail: kast@iesl.forth.gr

\end{document}